\documentclass[twocolumn,showpacs,preprintnumbers,amsmath,amssymb]{revtex4}

\usepackage{graphicx}
\usepackage{dcolumn}
\usepackage{bm}
\usepackage{multirow}

\begin{document}


\title{Precision spectroscopy of pionic 1s states of Sn nuclei and
evidence for partial restoration of chiral symmetry in the nuclear medium}%

\author{K. Suzuki$^a$, M.~Fujita$^b$,
H.~Geissel$^c$,
H.~Gilg$^d$,
A.~Gillitzer$^e$,
R.S.~Hayano$^a$,
S.~Hirenzaki$^b$,
K.~Itahashi$^f$,
M.~Iwasaki$^f$,
P.~Kienle$^{d,g}$,
M.~Matos$^c$,
G.~M\"unzenberg$^c$,
T.~Ohtsubo$^h$,
M.~Sato$^i$,
M.~Shindo$^a$,
T.~Suzuki$^a$,
H.~Weick$^c$,
M.~Winkler$^c$,
T.~Yamazaki$^j$,
T.~Yoneyama$^i$}
\address{$^a$Department of Physics, School of Science, University of
Tokyo, 7-3-1 Hongo, Bunkyo-ku, Tokyo 113-0033, Japan,}
\address{$^b$Department of Physics, Nara Women's University, Kita-Uoya
Nishimachi,
Nara 630-8506, Japan,}
\address{$^c$Gesellschaft f\"ur Schwerionenforschung,
D-64291 Darmstadt, Germany,}
\address{$^d$Physik-Department, Technische Universit\"at M\"unchen,
D-85748 Garching, Germany,}
\address{$^e$Institut f\"ur Kernphysik, Forschungszentrum J\"ulich,
D-52425 J\"ulich, Germany,}
\address{$^f$Muon Science Laboratory, RIKEN,  2-1 Hirosawa,
Wako-shi, Saitama, 351-0198 Japan,}
\address{$^g$Institute for Medium Energy Physics of the Austrian
Academy of Sciences, Boltzmanngasse 3, A-1090 Vienna, Austria,}
\address{$^h$Department of Physics, Niigata University,
2-8050, Igarashi, Niigata-shi, Niigata, 950-2181, Japan,}
\address{$^i$Department of Physics, Tokyo Institute
of Technology, 2-12-1 Ookayama, Meguro-ku, Tokyo 152-8551, Japan,}
\address{$^j$RI Beam Science Laboratory, RIKEN,  2-1 Hirosawa,
Wako-shi, Saitama, 351-0198 Japan}


\date{\today}

\begin{abstract}
Deeply bound 1s states of $\pi^-$ in $^{115,119,123}$Sn were preferentially
observed using the Sn($d$,$^3$He) pion-transfer reaction under the
recoil-free condition. The 1s binding energies and widths were precisely
determined, and were used to deduce the isovector
parameter of the s-wave pion-nucleus potential to be
$b_1 =-0.115\pm 0.007 ~m_{\pi}^{-1}$.
The observed enhancement of $|b_1|$
over the free $\pi N$ value ($b_1^{\rm free}/b_1 = 0.78 \pm 0.05$) indicates a
reduction of the chiral order parameter,
$f^{*}_{\pi} (\rho)^2/f_{\pi}^2 \approx 0.64$,
at the normal nuclear density, $\rho = \rho_0$.
\end{abstract}

\maketitle

Our work deals with characteristic properties of strongly
interacting bound sytems, hadrons, consisting of light quarks
($u$ and $d$). 
Their masses are nearly two orders of magnitude smaller
($m_u \simeq 5$MeV, $m_d \simeq 8$MeV) than typical
hadron masses of $\approx$ 1 GeV.
This extraordinary phenomenon is proposed to be produced by spontaneous
breaking of chiral symmetry for massless quarks subject to the strong
interaction~\cite{Nambu, Hatsuda94, Vogl91}.
It results in a ground state, the vacuum state, with a finite
expectation value of quark-antiquark pairs,
$\langle \bar{q}q \rangle _0 \approx -(250~{\rm MeV})^{3}$~\cite{GOR}.
In such a scenario the hadrons are considered as quasi particle
excitations of the condensate, $\langle \bar{q}q \rangle$, separated
by an energy gap of $\sim$ 1 GeV.
The lowest energy excitation modes of the
condensate, so called Nambu-Goldstone bosons, are identified as pions.
Their s-wave interaction with nucleons is predicted to
vanish in its isoscalar part, and determined by the pion decay
constant, $f_{\pi}^2$, in its isovector part~\cite{Tomozawa,Weinberg}
which is consistent
with recent experimental values~\cite{Schroeder99, Ericson2000}.
The $f_{\pi}^2$ is also the order parameter of
chiral symmetry breaking and directly connected to the magnitude of
$\langle \bar{q}q \rangle$ through the Gell-Mann-Oaks-Renner
relation~\cite{GOR}. 

We examine this chiral symmetry scenario by implanting a
$\pi^-$ in a nuclear medium of density $\rho$ \cite{Weise2000,
Kienle:01}, where a new vacuum state with a reduced condensate,
$\langle \bar{q}q \rangle _{\rho}$, is proposed to be created
\cite{Hatsuda94}.
This is the most basic case of a large effort to study the density and
temperature dependence of the $\langle \bar{q}q \rangle$ 
in relativistic heavy ion collisions.

The density dependence of the quark condensate is expressed in the
leading order by the pion-nucleon $\sigma$-term
($\sigma_N \approx 45~{\rm MeV}$) in the following form \cite{Drukarev}:
\begin{equation}\label{eq:condensate}
        \frac{\langle \bar{q}q
\rangle _{\rho}}{\langle \bar{q}q
\rangle _{0}} \approx 1 - \frac{\sigma_N}{m^2_\pi f_{\pi}^2}\rho,
\end{equation}
which yields a reduction of about 0.65 for the normal nuclear density,
$\rho = \rho_0 = 0.17~{\rm fm}^{-3}$.
Likewise, the pion decay constant in a medium
(identified as the  time component of the
axial current) is reduced as~\cite{Thorsson},
$\frac{f_{\pi}^* (\rho)^2}{f_{\pi}^2} \approx 1 - \alpha \rho$,
where the parameter $\alpha$ is predicted to be $\alpha \rho_0 \approx
0.45$ from the chiral dynamics \cite{Meissner2002}.
This reduction of the ratio
is associated with the free and the in-medium
isovector $\pi N$ 
scattering length
($b_1^{\rm free}$ and $b_1^*$, respectively)
as~\cite{Kolomeitsev:02}
\begin{equation}\label{eq:b1}
R(\rho) = \frac{b_1^{\rm free}}{b_1^* (\rho)} \approx \frac{f_{\pi}^*
(\rho)^{2}}{f_{\pi}^{2}} \approx 1 - \alpha \rho.
\end{equation}

Our program aims at measuring
the isovector
$\pi N$ interaction parameter in the pion-nucleus potential
($b_1^* (\rho)$)~\cite{EE}
by studying deeply bound 1s states of $\pi^-$ in
heavy $N>Z$ nuclei \cite{Zphys,Yamazaki:98,PRC1,PRC2,PRL}.
Such states were predicted to be produced as discrete states by
nuclear reactions~\cite{Toki88,Toki89,Toki91,Hirenzaki91,Umemoto} and
to provide unique information on the s-wave interaction, whereas in
most pionic atoms the repulsive s-wave interaction is nearly
counterbalanced by the attractive p-wave interaction (see, for
instance, Batty {\it et al.} \cite{Batty97}).
In recent papers we showed that the density-dependent parameter
$b_1^*({\rho})$ can be well represented by a constant parameter
$b_1$~\cite{YH02}, and developed a method to deduce the $b_1$
parameter~\cite{Geissel:02}.
The data of the 1s $\pi^-$ state in $^{205}$Pb~\cite{PRL},
combined with information on 1s states of light pionic atoms, yielded
$b_1 = -0.116^{+0.015}_{-0.017}~m_{\pi}^{-1}$.
A similar conclusion was also obtained from global fits of pionic atom
data by Friedman, $b_1 = -0.122 \pm 0.004~m_{\pi}^{-1}$
\cite{Friedman:02a} and $-0.113 \pm 0.004~m_{\pi}^{-1}$
\cite{Friedman:02b}, although
some ambiguity arising from the p-wave part remains.
In order to determine $b_1$ more reliably and accurately, it
is essential to perform high-precision spectroscopy on deeply bound 1s
$\pi^-$ states in heavy nuclei.

In the present paper we report on systematic experimental studies of
1s $\pi^-$ states in a series of Sn isotopes, which were produced with
the Sn($d$,$^3$He) reactions.
One of the advantages of using Sn isotopes is that we can produce the
1s $\pi^-$ states as the most dominant quasi-substitutional states,
$(1s)_{\pi^-}(3s)_n^{-1}$, because of the presence of the 3s orbital
near the Fermi surface, as theoretically predicted~\cite{Umemoto}.
Another merit is to make use of isotopes over a wide range of
$(N-Z)/A$ to test the isospin dependence~\cite{Kienle:01}.

We used a deuteron beam from the heavy-ion
synchrotron SIS at GSI, Darmstadt, combined with  the Fragment
Separator as a high-resolution forward spectrometer \cite{PRC1}.
The present experiment had many new features and improvements.
We chose the exact recoilless condition to suppress minor states other
than the enhanced 1s $\pi^-$ states with quasi-substitutional
1s $\pi^-$ states,
with a deuteron beam of a small momentum spread 
and an accurately measured energy of 503.388$\pm$ 0.100 MeV.
Its intensity was $1.5\times 10^{11}$ deuterons per spill.
The beam was focused on a narrow spot with a width of about 1 mm,
hitting a strip target with a width of 1.5 mm and a thickness of
20 mg/cm$^2$. Enriched Sn isotopes
with $A$ = 116 (98.0 \%), 120 (99.2 \%) and 124 (96.6 \%) were used as
targets.

\begin{figure}[htb]
\includegraphics[height=13.5cm]{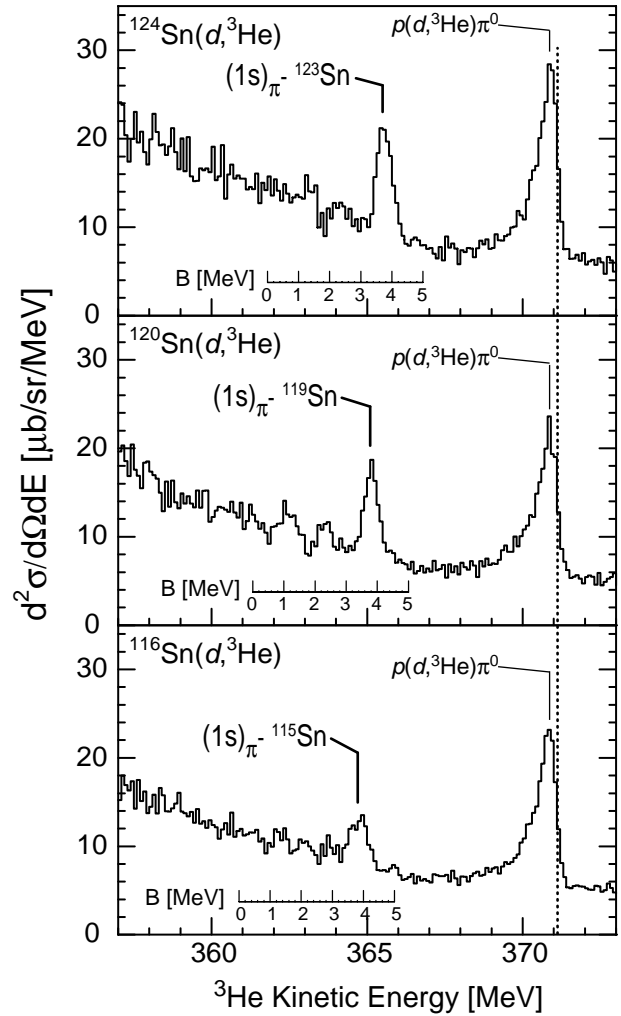}
\caption{\label{fig:spectra} Double differential cross sections versus the
        $^3$He kinetic energy of the $^{124, 120, 116}$Sn($d$,{\rm $^3$He)}
        reactions measured at the incident deuteron energy of 503.388\ MeV.
	The scales of the $\pi^-$ binding energies in $^{123, 119,
        115}$Sn are also indicated.}
\end{figure}

The strategy to arrive at high accuracy in determining the binding
energies and widths of the 1s states was the following.
First, we used the non-pionic 
$^{A}$Sn(d,$^3$He)$^{A-1}$In(g.s.) ($A = 112, 116, 120, 124$)
reactions to calibrate the spectrometer momentum (with respect to the
incident beam momentum) and dispersion to achieve an accuracy of 6
keV.  The overall energy resolution, which was crucial for measuring
the widths of the 1s $\pi^-$ states, was also determined from these
spectra to be $394 \pm 33$ keV (FWHM). Secondly, we put a thin mylar 
layer on the down-stream surface of each Sn target to measure the
$^3$He peak from the $p$($d$,$^3$He)$\pi^0$ reaction as a built-in
indicator of absolute calibration. We observed spectra, 
$d^2\sigma/dE/d\Omega$,  on  mylar-covered
$^{116}$Sn, $^{120}$Sn, $^{124}$Sn targets as function of the $^3$He
kinetic energy, as shown in Fig.~\ref{fig:spectra}.

The skewed peaks at around 371 MeV arise from $p$($d$,$^3$He)$\pi^0$.
Their kinematically broadened shapes were well explained by a
simulation, which took into account the beam-energy profile at
the thin mylar layer, the angular distribution, the spectrometer
acceptance
($\Delta \theta_x \approx \pm 15 ~{\rm mr}$ and $\Delta \theta_y
\approx \pm 10 ~{\rm mr}$) and the kinematical 
energy shifts of the $^3$He particles. It turned out that the energy
at half maximum of the higher-energy
edge of the peak (shown as vertical dotted lines in Fig.~\ref{fig:spectra})
represents the kinetic energy of $^3$He in the ideal case of 0-degree
emission from the $p$($d$,$^3$He)$\pi^0$ reaction, and
serves as an absolute calibration point with an accuracy of $\pm$ 7
keV. We adopted this ``edge" calibration for our final results, since
this long term calibration was pertinent to all runs of the three
targets. It  was found to agree within 13 keV with the short-term
calibration using the Sn($d, ^3$He)In(g.s.) reactions.

\begin{table*}[htb]
\caption{\label{tab:BE} Observed binding energies ($B_{1s}$) and widths
($\varGamma_{1s}$) of the 1s $\pi^-$ states in $^{115, 119, 123}$Sn isotopes.}
\begin{ruledtabular}
\begin{tabular}{p{2.5cm}p{1.5cm}p{1.5cm}p{1.5cm}p{2.0cm}p{1.5cm}p{1.5cm}p{1.5cm}p{1.5cm}}
\multirow{2}{2cm}{Isotope}&
\multirow{2}{4cm}{B$_{1s}~{\rm [MeV]}$} & &$\Delta B_{1s}~{\rm [MeV]}$ &&
\multirow{2}{4cm}{$\varGamma_{1s}~{\rm [MeV]}$} & &$\Delta \varGamma_{1s}~{\rm
[MeV]}$ &\\
\cline{3-5} \cline{7-9}
   & & stat. & syst. & total &       & stat. & syst. & total\\ \hline
$^{115}$Sn & 3.906 & $\pm 0.021$ & $\pm 0.012$ & $\pm 0.024$ &
               0.441 & $\pm 0.068$ & $\pm 0.054$ & $\pm 0.087$\\
$^{119}$Sn & 3.820 & $\pm 0.013$ & $\pm 0.012$ & $\pm 0.018$ &
	     0.326 & $\pm 0.047$ & $\pm 0.065$ & $\pm 0.080$\\
$^{123}$Sn & 3.744 & $\pm 0.013$ & $\pm 0.012$ & $\pm 0.018$ &
	     0.341 & $\pm 0.036$ & $\pm 0.063$ & $\pm 0.072$\\
\end{tabular}
\end{ruledtabular}
\end{table*}

\begin{figure}[htb]
\includegraphics[height=7.5cm]{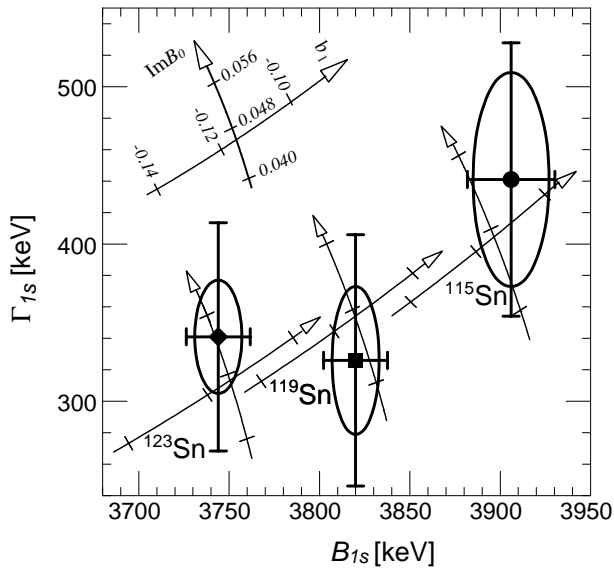}
\caption{\label{fig:BvsG}
 Experimental values of \{$B_{1s}$,$\varGamma_{1s}$\}
 (elliptic circles for statistical errors; horizontal and vertical
 bars for total errors) for $\pi^-$ in $^{115, 119, 123}$Sn, shown
 together with theoretical relations with \{$b_1$, Im$B_0$\} for each
 isotope.
}
\end{figure}

In each spectrum of Fig.~\ref{fig:spectra} one distinct peak at
around 365 MeV was observed, which was assigned to a dominant
configuration of $(1s)_{\pi}(3s)_n^{-1}$.
The overall spectrum shapes for the three Sn targets were found to
be in good agreement with the predicted ones~\cite{Umemoto}.
The spectra were decomposed according to the theoretical
prescription of Ref.~\cite{Umemoto},
from which we could precisely determine the 1s binding energies
($B_{1s}$) and widths ($\varGamma_{1s}$).
The results are summarized in Table~\ref{tab:BE} and in
Fig.~\ref{fig:BvsG}.
The quoted errors are statistical, systematic (arising from the
calibration) and combined errors.
In this study isotope shifts of deeply bound pionic 1s states in heavy
nuclei are seen for the first time.

The most serious problem in our analysis is the relatively poor knowledge
concerning the neutron distribution $\rho_n (r)$ in Sn isotopes,
whereas the proton distribution
$\rho_p (r)$ is well known \cite{Fricke}. We take the
following form for the difference between the neutron and proton rms radii,
which was derived based on experimental data of antiprotonic atoms of
Sn isotopes~\cite{Trzcinska} as well as of many other nuclei
       $ \Delta r_{np} =
     (1.01 \pm 0.15) (N-Z)/A + (-0.04 \pm 0.03) ~{\rm fm}.$
In converting $\Delta r_{np}$ into the diffuseness parameters ($a_p, a_n$) 
and the half-density radii ($c_p, c_n$) in the two-parameter
Fermi distribution we took half way between
the ``halo" type ($c_p = c_n, a_p < a_n$) and
the ``skin" type ($c_p < c_n, a_p = a_n$)\cite{Trzcinska},
in accordance with a proton elastic scattering result in Pb isotopes
(see Ref.~\cite{Geissel:02}).
The adopted parameters \{$c_p$, $a_p$, $c_n$, $a_n$\} in fm are:
\{5.405, 0.523, 5.473, 0.552\} for $^{115}$Sn,
\{5.442, 0.523, 5.532, 0.560\} for $^{119}$Sn, and
\{5.484, 0.523, 5.594, 0.569\} for $^{123}$Sn.

The influence of $\Delta{}r_{np}$ on the extraction of optical
potential parameters in global fits of pionic atom data was recently
discussed in detail by Friedman and Gal~\cite{FG03}
based on old theoretical values of $\Delta{}r_{np}$~\cite{RingOld},
which are considerably different from
the empirical ones~\cite{Trzcinska}we use.
These values are supported by a new proton-scattering result on Sn
isotopes~\cite{RCNP} and also by a new RHB calculations~\cite{NVFR02}.
In our view the use, in our analysis, of the experimental data
available now resolves the concerns of Ref.~\cite{FG03} with regard to
our results.

\begin{figure}[htb]
\includegraphics[width=8.2cm]{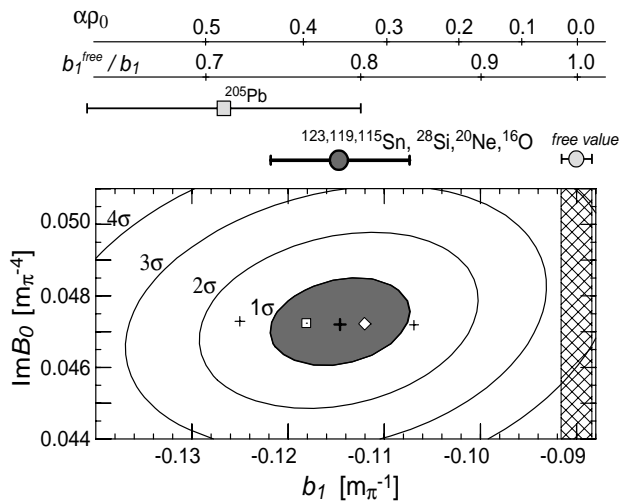}
\caption{\label{fig:contour}
Likelihood contours in the \{$b_1$, Im$B_0$\} plane from the simultaneous 
fitting of \{$B_{1s}$, $\varGamma_{1s}$\} of the 1s pionic states in
the three Sn isotopes and three light symmetric nuclei.
   The previous $^{205}$Pb data reanalyzed with
   Im$B_0$ ($= 0.046$) is shown for comparison.}
\end{figure}

  We made  simultaneous fitting of $B_{1s}$ and $\varGamma_{1s}$ of
the three Sn isotopes together with those of symmetric light nuclei,
leaving $b_0$, $b_1$, Re$B_0$ and Im$B_0$ as free parameters. 
 Since the 1s binding energy is insensitive to the p-wave potential,
 we could safely fix the p-wave parameters to the known values from
 pionic atom data~\cite{Batty97}. Hereafter, the units of
 $m_{\pi}^{-1}$ for $b_0, b_1$ and of $m_{\pi}^{-4}$ for $B_0$ will be
 omitted.
 We chose
$^{16}$O, $^{20}$Ne and $^{28}$Si for which the nuclear parameters are 
well known and the condition $\rho_p (r) = \rho_n (r)$ is clearly
fulfilled. 
Thus the obtained values are:
$b_0 = -0.0233 \pm 0.0038, b_1 = -0.1149 \pm 0.0074,
{\rm Re}B_0 = -0.019 \pm 0.017, {\rm Im}B_0 = 0.0472 \pm 0.0013.$
The errors include both statistical and systematic uncertainties.
In Fig.~\ref{fig:BvsG} we show the obtained relations of \{$b_1$, Im$B_0$\} 
mapped on \{$B_{1s}$, $\varGamma_{1s}$\} for each Sn isotope. The
intersections of the tilted axes are chosen to the best-fit values of
\{$b_1$, Im$B_0$\}. We also show likelihood contours in the plane of
\{$b_1$, Im$B_0$\} in Fig.~\ref{fig:contour}.

The smallness of Re$B_0$ ($-0.019 \pm 0.017$) supports the
value ($-0.038 \pm 0.025$) claimed from the requirement that the
effective $b_0$ consists of the free $\pi N$ value and the
double-scattering term \cite{Geissel:02}.
The combined isoscalar parameter,
$b_0 + 0.215\, {\rm Re}B_0 = -0.0274 \pm 0.0002$, is in good 
agreement with that obtained in the analysis of the $^{205}$Pb
\cite{Geissel:02}.
The imaginary part, Im$B_0 = 0.0472 \pm 0.0013$, is consistent with the 
global-fit value of
$0.055 \pm 0.003$ by Batty {\it et al.} \cite{Batty97} and Friedman
\cite{Friedman:02a,Friedman:02b}, considering that they included the 
angle-transformation (AT) term, which causes an appreciable 
decrease in the width~\cite{YH02}.
In fact, the best-fit value in our analysis with the AT term 
included is: Im$B_0 = 0.058 \pm 0.003$.
These isoscalar parameters are in fine agreement with those determined
by using only the 1s states of light symmetric pionic atoms.
The addition of the Sn data consequently yields a precise value for
$b_1$.

Figure~\ref{fig:contour} also shows how the best-fit values move, if we 
choose the ``halo" type (open square) or the ``skin" type (open diamond) 
for $\rho_n (r) - \rho_p (r)$.
Furthermore, we find that the possible change of  $b_1$, when we
allowed an uncertainty of $\pm 0.04$ fm in $\Delta r_{np}$
\cite{Trzcinska}, would be around $\pm 0.009$,
as indicated by two crosses in the figure.

The magnitude of the observed $|b_1|$ is significantly enhanced over
the free $\pi N$ value, which translates into a reduction of
${f_\pi ^*}^2$ (Eq.~\ref{eq:b1}) as
\begin{equation}
R  = {b_1^{\rm free}}/{b_1} =0.78 \pm 0.05
\end{equation}
\begin{equation}
   \approx {b_1^{\rm free}}/{b_1^* (\rho_e)} 
    \approx {f_{\pi}^* (\rho_e)^2}/{f_{\pi}^2}
    \approx 1 - \alpha \rho_e,
\end{equation}
where we made use of the fact \cite{Geissel:02,YH02} 
that the solution with a local-density-dependent parameter,
$b_{1}^* (\rho) = b_1^{\rm free}/(1 - \alpha \rho(r))$,
is equivalent to that using a corresponding constant parameter 
$b_1 = b_1^{\rm free}/(1-\alpha \rho_e)$ 
with an effective density $\rho_e\approx 0.6 \rho_0$.

The above value hence implies that
the chiral order parameter, $f_{\pi}^* (\rho)^2$, would be reduced
by a factor of $\approx 0.64$
at the normal nuclear density $\rho = \rho_0$.
Using Eq.(\ref{eq:b1}) in the analysis, we
obtain an experimental value of $\alpha \rho_0 \approx 0.36$,
as shown in Fig.~\ref{fig:contour},
which is close to the value 0.45 predicted by chiral perturbation
theory~\cite{Meissner2002}.
If a theoretical value, $m_{\pi}^* \approx m_{\pi} + 3~{\rm MeV}$
(averaged over $\pi^+$ and $\pi^-$ \cite{Meissner2002}), is inserted
into  an in-medium Gell-Mann-Oakes-Renner relation
\cite{Hatsuda94, Thorsson},
$\langle \bar{q}q \rangle _{\rho_0}/\langle \bar{q}q \rangle _{0}$
will be $(m_{\pi}^*/m_{\pi})^2 \times (1 - \alpha \rho_0)
\approx 0.67$,
which is in good agreement with the value of 0.65, as predicted by using
Eq.(\ref{eq:condensate}). We have thus found clear evidence for the partial 
restoration of chiral symmetry, probed by {\it well-defined} pionic states 
in a {\it well-defined} nuclear density.
 
The authors would like to thank the staff
of GSI for the continuous efforts to provide superb experimental
conditions and Professors H. Toki, W. Weise and T. Hatsuda
for stimulating and encouraging discussions. This work is supported by
Grants-in-Aid for
Scientific Research of Monbukagakusho (Japan) and Japan Society for the
Promotion of Science, and by the Bundesministerium f\"ur
Bildung, Wissenschaft, Forschung und Technologie (Germany) and the
Gesellschaft f\"ur Schwerionenforschung Darmstadt.

\end{document}